# Pressure-driven evolution of crystal and molecular structures reshapes excited-state competition in coronene

Wenju Zhou[1,†], Shuhua Yuan[2,†], Dominique Laniel[3], Yoshifumi Hashikawa[4], Philip Dalladay-Simpson[1], Jun Deng[1], Jianbo Zhang[1], Xinyu Wang[1], Kejun Bu[1], Akun Liang[3], Simon Parsons[5], Dominik Daisenberger[6], Egor Koemets[6], Hirofumi Ishii[7], Weiwei Dong[8], Huiyang Gou[1], Qingyang Hu[1], Yang Ding[1], Martina Vrankić[9,*], and Takeshi Nakagawa[1,*,‡]

[1]Center for High Pressure Science and Technology Advanced Research, 100193 Beijing, China

[2]Longyan Fangmei Institute for Educational Development, Longyan 364000, Fujian, China

[3]Centre for Science at Extreme Conditions and School of Physics and Astronomy, University of Edinburgh, EH9 3FD Edinburgh, United Kingdom

[4]Institute for Chemical Research, Kyoto University, 611-0011 Kyoto, Japan

[5]Centre for Science at Extreme Conditions and School of Chemistry, University of Edinburgh, EH9 3FD Edinburgh, United Kingdom

[6]Harwell Science and Innovation Campus, Diamond Light Source, Didcot, Oxfordshire, UK

[7]National Synchrotron Radiation Research Center, Hsinchu 30076, Taiwan, R. O. C.

[8]Beijing Synchrotron Radiation Facility, Institute of High Energy Physics, Chinese Academy of Sciences, 100049 Beijing, China

[9]Division of Materials Physics, Ruder Bošković Institute, Bijenička 54, 10000 Zagreb, Croatia

**Abstract** Molecular nanocarbons exhibit unique optoelectronic properties and considered as promising candidates for organic device applications, however, understanding how molecular geometry and crystal packing jointly govern physical properties remains a fundamental challenge in π-conjugated molecular materials. Here, we demonstrate how these structural factors modulate the photoluminescence of coronene under high pressure. Combining high-pressure single-crystal X-ray diffraction, Raman spectroscopy, optical spectroscopy, and density functional theory calculations, we establish a relationship between distinct structural changes and the evolution of competing emissive states. Our results reveal that crystal polymorphism modifies the balance between free-exciton (FE) and self-trapped-exciton (STE) emission, while pressure-induced molecular bending is associated with a nonmonotonic evolution of STE emission. As intermolecular distances shorten, enhanced intermolecular electronic interactions give rise to excimer-related emission, introducing an additional emissive pathway. Although pressure-dependent optical properties of coronene have been linked to changes in crystal packing, our findings highlight that the optical response of a molecular solid is not governed by molecular deformation or crystal packing alone, but emerges from their interplay with intermolecular electronic coupling. By directly resolving molecular bending alongside changes in excited-state emission, this work provides insight into how distinct structural degrees of freedom can reshape the photophysical behavior of π-conjugated molecular solids.

**Keywords:** High-pressure chemistry, Polycyclic aromatic hydrocarbons, Molecular deformation, crystal phase transition, photoluminescence, Intermolecular interactions, structure-property relationships

*Correspondence: Martina.Vrankic@irb.hr; takeshi.nakagawa@hpstar.ac.cn

## 1. Introduction

Molecular arrangement in solid is a fundamental determinant of the electronic and optical properties of π-conjugated materials.[1-4] In molecular crystals, however, the molecular framework is often treated as a fixed structural unit, which changes in physical properties primarily attributed to variations in crystal packing or intermolecular interactions.[5,6] This is particularly intriguing for polycyclic aromatic hydrocarbons (PAHs), whose molecular structures are extremely stable and delocalized aromatic π-electron systems properties that are highly sensitive to molecular packing.[7,8] Deviations from planar π-conjugated molecule into distorted configurations, such as bent, curved, bowled, twisted, or warped, is expected to modify π-electron delocalization and excited-state relaxation.[9-13] Molecular distortion therefore represents an additional structure degree of freedom through which the properties of π-conjugated molecular materials may be modified. Understanding how molecular deformations evolve and govern functional properties is particularly important for stimuli-responsive molecular solids.

Realizing a high-pressure environment by the aid of a diamond anvil cell (DAC) provides a distinctive approach to investigating direct relationship between structure transitions and property responses.[14,15] Hydrostatic pressure can modify molecular conformation and crystal packing within the same material without changing the chemical composition of the molecular framework. In particular, compression on π-conjugated systems is expected to bring π-surface contacts into a region inaccessible under ambient conditions while continuously altering intermolecular distances and orientations, thereby changing electronic coupling between neighboring molecules.[16,17] Consequently, optical responses under pressure can reflect competing contributions from intramolecular geometry and intermolecular interactions.[18,19] Establishing a direct relationship between molecular deformation and optical response remains challenging, as experimental evidence directly linking progressive molecular deformation and excited-state behavior in stimuli-responsive molecular materials is still lacking.

PAHs provide useful model systems for exploring this challenge because their fused aromatic frameworks offer well-defined molecular geometries and extended π-conjugated structures. Indeed, recent high-pressure studies have begun to reveal the structural flexibility of non-planar π-conjugated molecular systems, reporting compression of all unit cell axes, optimization of dihedral angles, and modulation of torsion angle contribute to their optical behavior.[20-22] Contrary to those non-planar systems, a disk shaped molecular nanocarbons coronene ($C_{24}H_{12}$) offers a compelling platform for investigating the deformation of an initially planar molecular framework. At ambient conditions, coronene crystallizes in a monoclinic γ-herringbone structure (space group $P2_1/n$), stabilized by a delicate balance of dispersion forces, C-H···π contacts, and π···π interactions.[23] Its photoluminescence comprises distinct free-exciton (FE) and self-trapped-exciton (STE) emission pathways, providing a sensitive probe of changes in the excited-state landscape.[24]

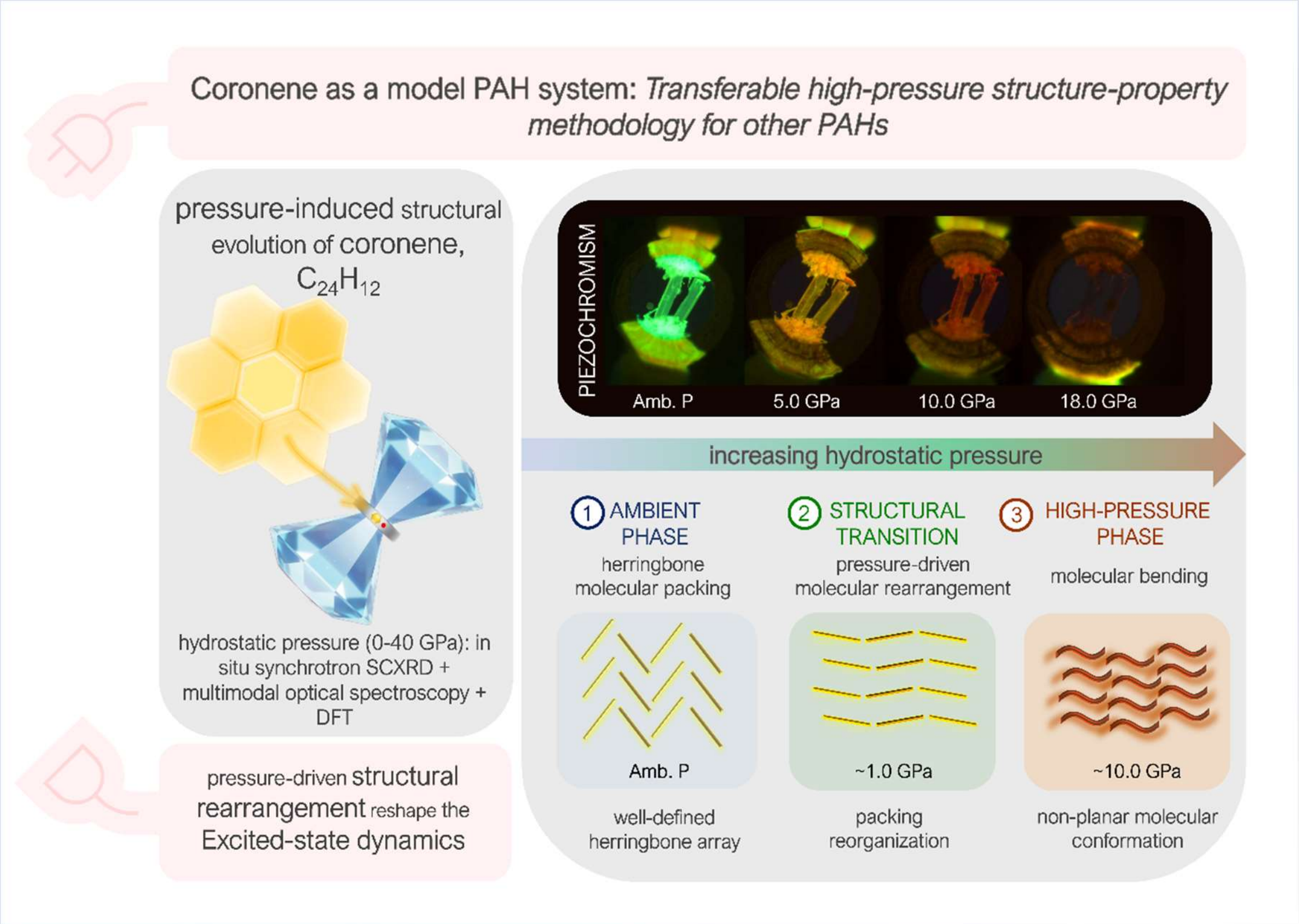

**Figure 1.** Molecular structure of coronene, DAC setup, and the interplay between pressure-induced structural changes to optical property.

Here, we investigate the pressure-induced solid-state phase transition, molecular deformation and optical response of crystalline coronene by combining high-pressure single-crystal X-ray diffraction (SC-XRD), Raman spectroscopy, optical absorption and photoluminescence measurements, and density functional theory (DFT) calculations. SC-XRD reveals an initial transition from the γ- to β-polymorph, followed by progressive bending deformation of the coronene framework with further compression. **(Figure 1)** Raman spectroscopy provides complementary vibrational evidence for changes in molecular distortion, while optical measurements reveal a strongly nonmonotonic photoluminescence response, including changes in the relative contributions of FE and STE emission, subsequent emission enhancement and quenching at higher pressure. DFT analyses provide additional insight into changes in aromaticity-related structural descriptors, relative structural stability, and intermolecular electronic interactions. This approach disentangles the effects of a pressure-induced polymorphic transformation from those of subsequent continuous molecular deformation, providing a direct experimental framework for elucidating how molecular bending and intermolecular coupling reshape the excited-state landscape. Our findings establish hydrostatic pressure as a precise tool for controlling molecular packing, intermolecular distances, and molecular deformation, opening a route to previously inaccessible functional polymorphs across a broader range of stimuli responsive molecular nanocarbons.

## 2. Results and Discussion

### 2.1. Evolution of crystal and molecular structures

Coronene powder was purchased from TCI Chemicals (>95% purity) and further purified by sublimation. Long, needle-shaped single crystals were obtained by dissolving the purified coronene in anhydrous toluene at 370 K under stirring, followed by cooling to room temperature at a rate of 1 K/min. Pressure-dependent synchrotron powder X-ray diffraction (P-XRD) measurements up to 18.8 GPa were first performed, revealing a structural phase transition below 1.0 GPa (Table S1; unit cell parameters were obtained by sequential Le Bail refinement of the P-XRD data using the converged model at each pressure point as the starting point for the subsequent dataset). To gain further insight into the solid-state structure of coronene, we subsequently performed SC-XRD measurements **(Figure 2)**. Diffraction data at ambient conditions and under high pressure (0.0001-19.5 GPa, at room temperature) were collected using a laboratory diffractometer and synchrotron beamlines, respectively.

At ambient conditions, coronene crystallizes in a monoclinic γ-herringbone structure ($P2_1/n$, γ-$C_{24}H_{12}$), with lattice parameters of $a$ = 10.085(1) Å, $b$ = 4.6907(3) Å, $c$ = 15.66(1) Å, and $\beta$ = 105.99(1)°, and a unit-cell volume of $V$ = 712.1(1) Å$^3$. The interplanar distance between adjacent parallel molecules along the $b$-axis is 3.454(3) Å. The projected π-overlap area between two parallel coronene molecules is 15.29 Å$^2$, corresponding to 42.7% of the total molecular π-surface area (36.0 Å$^2$). These structural parameters are consistent with previously reported results.[25,26] Detailed crystallographic information is provided in Tables S2–S6.

At 1.0 GPa, the crystal structure was refined in the monoclinic space group $P2_1/n$, with lattice parameters of $a$ = 10.206(1) Å, $b$ = 3.6914(5) Å, $c$ = 17.00(3) Å, and $\beta$ = 95.83(4)°, and a unit-cell volume of $V$ = 637.3(1) Å$^3$. The structural differences between the ambient-pressure and high-pressure phases are characterized by a reduction in the interplanar distance from 3.454(3) Å to 3.368(1) Å, a change in the herringbone angle from 85.14(1)° to 48.3(1)°, and an increase in the π-overlap area between two parallel coronene molecules from 15.29 to 25.14 Å$^2$ (68.8% of the molecular π-surface area; Figure 2d). Crystallographically, the structure obtained at 1.0 GPa corresponds to the β-polymorph of coronene (β-$C_{24}H_{12}$), previously observed at 80 K and ambient pressure.[6]

We further collected SC-XRD data for β-$C_{24}H_{12}$, up to 19.5 GPa (Figure S1, Table S7), with the pressure-dependent lattice parameters summarized in Figure 2c and Table S8. Above 1.0 GPa, the crystal exhibited a pronounced anisotropic response to compression. By 19.5 GPa, the lattice parameter $b$ had decreased by 16.3% relative to its value at 1.0 GPa, whereas $a$ and $c$ had decreased by only 7.5% and 6.5%, respectively. This anisotropy indicates that the crystallographic direction associated with the $b$-axis is the most compressible, consistent with the direction perpendicular to the π-stacking planes. These observations suggest that the intermolecular packing, particularly the π···π stacking arrangement, plays an important role in governing the mechanical response of coronene under pressure. The pressure-volume data for β-$C_{24}H_{12}$ obtained by SC-XRD were fitted using a third-order Birch-Murnaghan equation of state (EoS) implemented in the EoSFIT software.[27] The resulting bulk modulus ($K_0$) was 8.1(6) GPa, with its pressure derivative ($K'_0$) of 9.3(8) and a fitted reference volume ($V_0$) of 320.1(6) Å$^3$. These results suggest that coronene is a relatively soft and compressible molecular solid. The unit-cell volume decreased by 27.6% at 19.5 GPa relative to that at 1.0 GPa (Figure 2b). The bulk modulus is comparable to those reported for PAHs, including naphthalene ($C_{10}H_8$, $K_0$ = 8.4(10) GPa), anthracene ($C_{14}H_{10}$, $K_0$ = 8.4(5) GPa),[28] and dicoronylene ($C_{48}H_{20}$, $K_0$ = 9.1(2) GPa).[29]

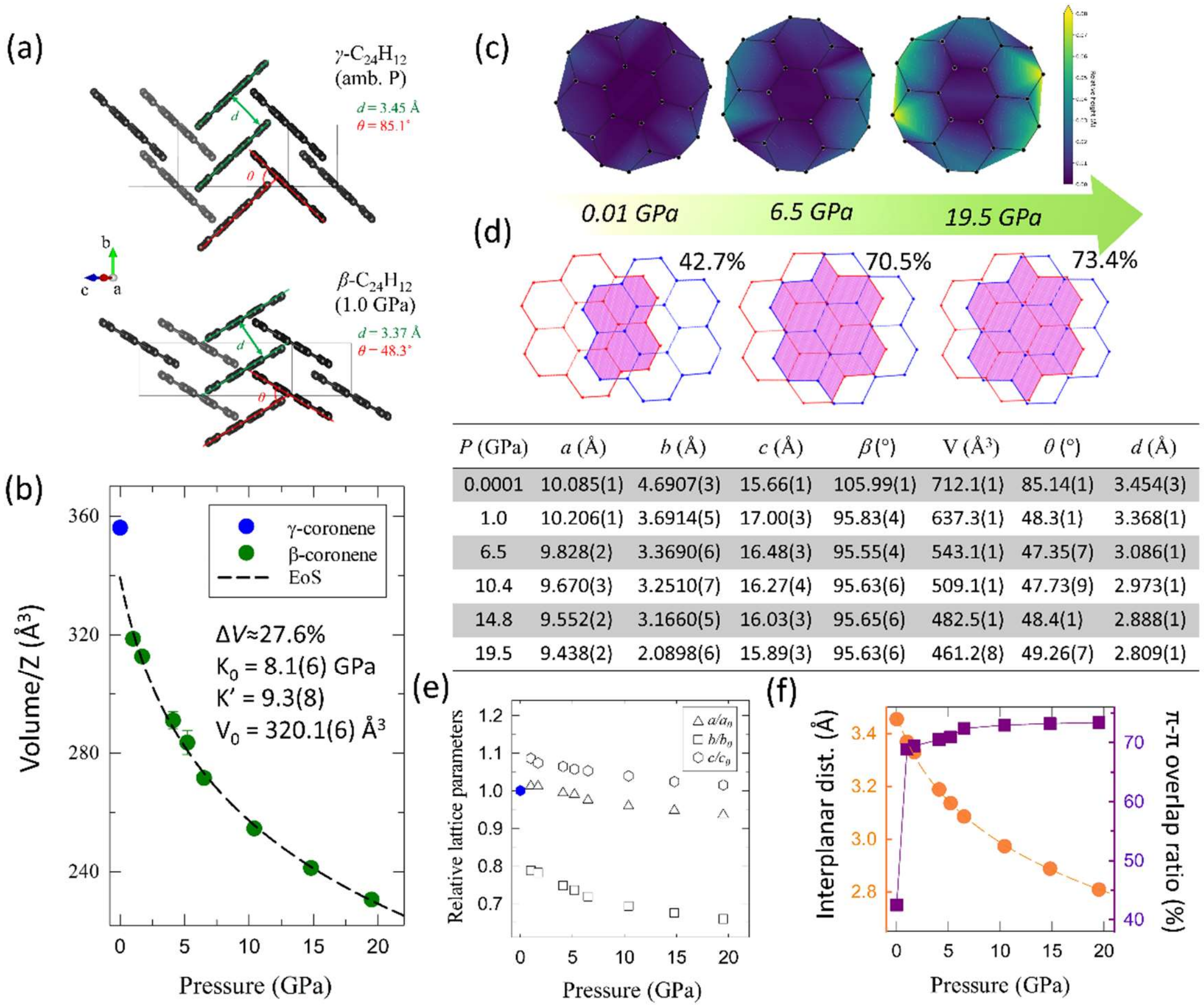

| $P$ (GPa) | $a$ (Å) | $b$ (Å) | $c$ (Å) | $\beta$ (°) | V (Å$^3$) | $\theta$ (°) | $d$ (Å) |
|---|---|---|---|---|---|---|---|
| 0.0001 | 10.085(1) | 4.6907(3) | 15.66(1) | 105.99(1) | 712.1(1) | 85.14(1) | 3.454(3) |
| 1.0 | 10.206(1) | 3.6914(5) | 17.00(3) | 95.83(4) | 637.3(1) | 48.3(1) | 3.368(1) |
| 6.5 | 9.828(2) | 3.3690(6) | 16.48(3) | 95.55(4) | 543.1(1) | 47.35(7) | 3.086(1) |
| 10.4 | 9.670(3) | 3.2510(7) | 16.27(4) | 95.63(6) | 509.1(1) | 47.73(9) | 2.973(1) |
| 14.8 | 9.552(2) | 3.1660(5) | 16.03(3) | 95.65(6) | 482.5(1) | 48.4(1) | 2.888(1) |
| 19.5 | 9.438(2) | 2.0898(6) | 15.89(3) | 95.63(6) | 461.2(8) | 49.26(7) | 2.809(1) |

**Figure 2. Pressure-induced structural evolution of coronene.** (a) Crystal structure of coronene at ambient conditions and 1.0 GPa. (b) Pressure dependence of the unit-cell volume per formula unit (V/Z) of γ-$C_{24}H_{12}$ (blue circle) and β-$C_{24}H_{12}$ (green circles), determined by SC-XRD. The solid line represents a fit of the third-order Birch-Murnaghan equation of state (EoS) to the β-$C_{24}H_{12}$ data. (the error bars are, in most cases, smaller than the data points) (c) Molecular surface maps and maximum height variations at selected pressures. (d) Maps of the π-π overlap between two parallel coronene molecules at 0.0001, 6.5 and 19.5 GPa. (e) Relative changes in the lattice parameters, referenced to their ambient condition values. (f) Pressure dependence of the interplanar distance and π-π overlap ratio.

To characterize the intermolecular contacts quantitatively, Hirshfeld surface analysis was performed using the SC-XRD structures.[30] The shape-index-mapped surfaces exhibit alternating red and blue triangular regions associated with π···π contacts, as well as concave red regions associated with C–H···π interactions. Compared with γ-$C_{24}H_{12}$, the alternating red and blue triangles on the Hirshfeld surface of β-$C_{24}H_{12}$ at 1.0 GPa become more equilateral, indicating a substantial change in the geometry of the π···π contacts (Figures S2–S5).

In addition to changes in crystal packing, SC-XRD revealed pressure-induced molecular deformation of the coronene molecules, which adopted a shallow chair-like conformation associated with enhanced vdW contacts near opposite ends of the molecular surface (Figure S6). The pressure dependence of the maximum molecular height reveals two distinct regimes of structural deformation. Below 5 GPa, the maximum height remains nearly constant at approximately 0.02 Å, comparable to or smaller than the average positional uncertainty (±1σ) of the carbon atoms. This indicates that the molecule remains essentially planar within the resolution of the structural refinement. Between 5 and 15 GPa, the maximum height increases progressively from approximately 0.025 to 0.08 Å, although the deviation remains relatively small compared with the positional uncertainties (less than 2σ) until pressures approach 15 GPa. Above 15 GPa, the maximum height stabilizes at approximately 0.08 Å, exceeding the average positional uncertainty. (Figure S8) These results suggest that measurable deviation from planarity develops at high pressures, although the magnitude of the distortion remains small. To quantify this deformation, we calculated the angle between two mean planes defined by distinct sets of carbon atoms. The resulting bending angle, $\theta_{bent}$, increased progressively above 5.2 GPa, from 1.3(4)° to 3.19(13)° at 19.5 GPa. The evolution of molecular curvature, where we quantified the maximum molecular height as the greatest perpendicular distance of the carbon atoms from the mean molecular plane, with pressure is illustrated in Figure 2c. The β-polymorph was retained up to 19.5 GPa, with no further crystallographic phase transition detected. At pressures beyond this limit, the diffraction reflections of coronene disappeared, most likely because of pressure-induced loss of crystallinity.

To further investigate the pressure-induced structural transitions at the crystal and molecular levels revealed by SC-XRD, we examined the pressure dependence of the inter- and intramolecular Raman modes. According to group theory, crystalline coronene possesses 216 vibrational modes, represented by $54A_g + 54A_u + 54B_g + 54B_u$. Of these, $54A_g + 54B_g$ modes are Raman-active, $53A_u + 52B_u$ modes are infrared-active, and the remaining $A_u + 2B_u$ modes correspond to acoustic vibrations (Figure S10, Table S10). Raman spectra were acquired using a 660 nm laser over the frequency range of 20-3300 $cm^{-1}$.

The pressure dependence of the lattice vibrational modes is shown in the left panel of **Figure 3**. In the low-frequency region, two lattice modes were observed at 60.5 $cm^{-1}$ and 89.6 $cm^{-1}$, corresponding to the two out-of-phase $A_g$ lattice modes of the stable γ-polymorph.[31] At 0.26 GPa, an abrupt change in the Raman spectrum indicated a first-order phase transition, characterized by the appearance of an additional $A_g$ mode at 42.6 $cm^{-1}$ and in-phase $B_g$ lattice modes at 17.9 and 76.8 $cm^{-1}$. These changes in the low-frequency region indicate a pressure-induced phase transition leading to a new polymorph. The newly observed spectral features closely match those previously reported for coronene at 160 K and ambient pressure, where they were assigned to the β-polymorph.[31,32] This assignment is consistent with that identified by SC-XRD. Upon further compression, all lattice modes progressively shifted to higher frequencies, with different pressure coefficients, leading to the resolution of two previously overlapping $A_g$ and $B_g$ modes near 85.0 $cm^{-1}$ above 0.86 GPa.

The pressure dependence of the intramolecular Raman modes is shown in the center and right panels of Figure 3. Of particular interest is a Raman feature near 850 $cm^{-1}$, assigned to an out-of-plane vibrational mode that is sensitive to deviations from molecular planarity.[33] Under ambient conditions, this mode appears as a very weak band at 853 $cm^{-1}$, with an integrated area of 270.4 counts·$cm^{-1}$ and a signal-to-noise ratio (SNR) of 5.3, and remains weak below 1.0 GPa. The initial enhancement was observed at 1.2 GPa, where its integrated area increases to 2215.5 counts·$cm^{-1}$, accompanied by an increase in SNR to 16.8, followed by a moderate increase in intensity up to approximately 4 GPa.

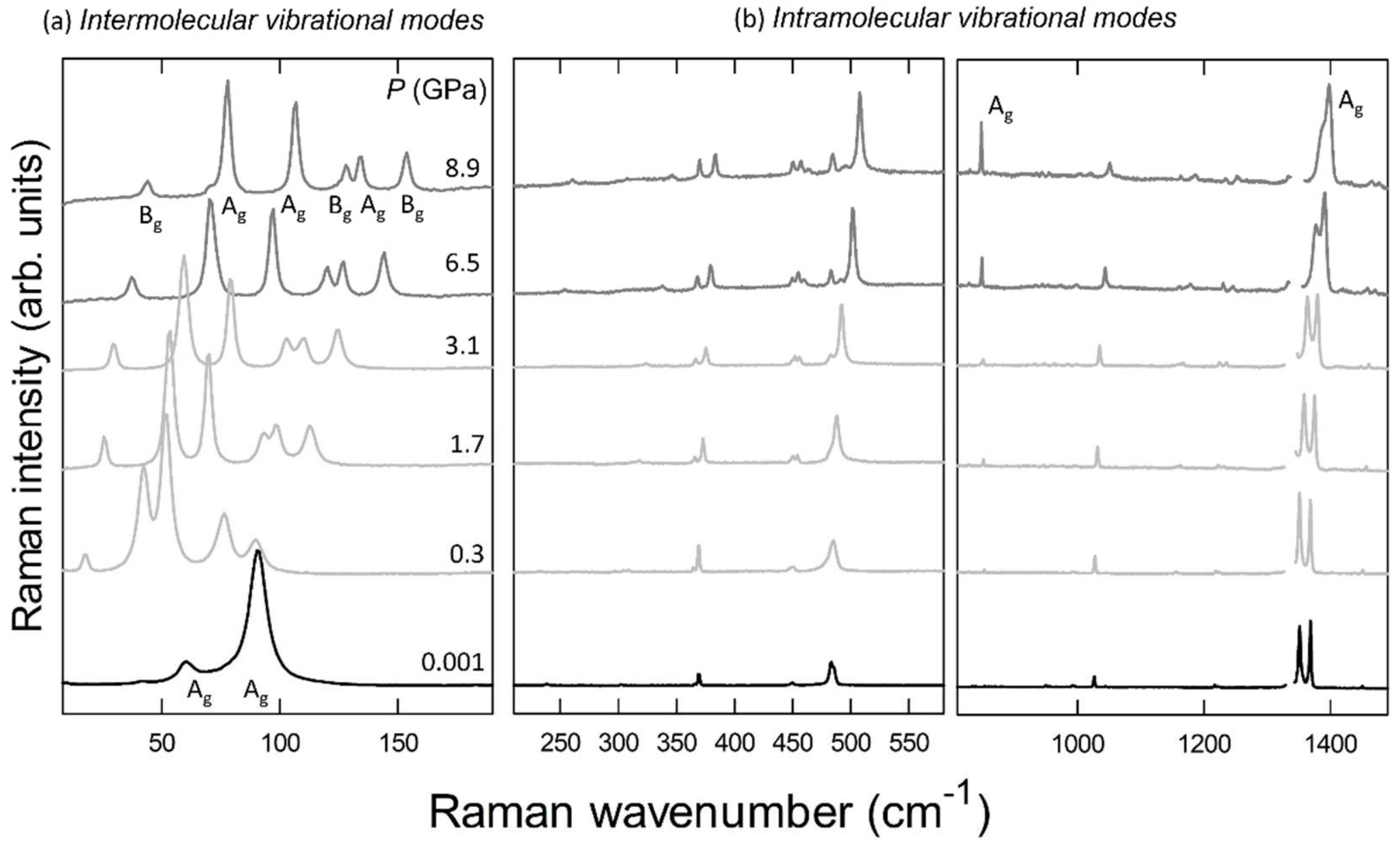

**Figure 3. High pressure Raman spectroscopy of coronene**. Pressure-dependent Raman spectra of a coronene single-crystal recorded during compression using 660nm excitation laser, showing (a) low-frequency intermolecular lattice modes and (b) intramolecular vibrational modes. Spectra of the γ- and β-polymorphs are shown in black and grey, respectively.

A marked enhancement occurs above approximately 5.0 GPa, where the integrated area increases by a factor of 4.5 and the SNR reaches 40. The band intensity continues to increase up to 8.9 GPa, at which the SNR reached approximately 84. This pronounced enhancement coincides with the substantial increase in the molecular bent angle determined by SC-XRD, suggesting a close relationship between the evolution of the out-of-plane vibrational response and pressure-induced molecular deformation. At higher pressures, however, a broad and intense fluorescence background spanning the entire measured spectral region obscures the Raman features, preventing their reliable identification. Upon decompression from 8.9 GPa, the Raman modes characteristic of the γ-polymorph re-emerged, although a small hysteresis was observed, with features of β-polymorph persisting down to 0.18 GPa(Figure S11). The γ-polymorph spectrum was fully recovered at 0.1 GPa. This hysteretic behavior indicates a kinetic barrier to the reverse phase transition and is consistent with metastable persistence of the β-polymorph during decompression.

## 2.2. Pressure-driven nonmonotonic photoluminescence response

Having established the pressure-induced evolution of the crystal structure and the accompanying changes in the intramolecular vibrational response, we next investigated how these structural modifications are reflected in the optical properties of coronene. In particular, the coexistence of crystal packing and progressive molecular deformation raises the question of how these distinct changes influence the excited-state landscape. Pressure-dependent optical images, photoluminescence (PL) microphotographs, ultraviolet-visible (UV-vis) absorption, and PL spectroscopies were collected from the same coronene single crystal loaded into a DAC. The results are summarized in Figure 4.

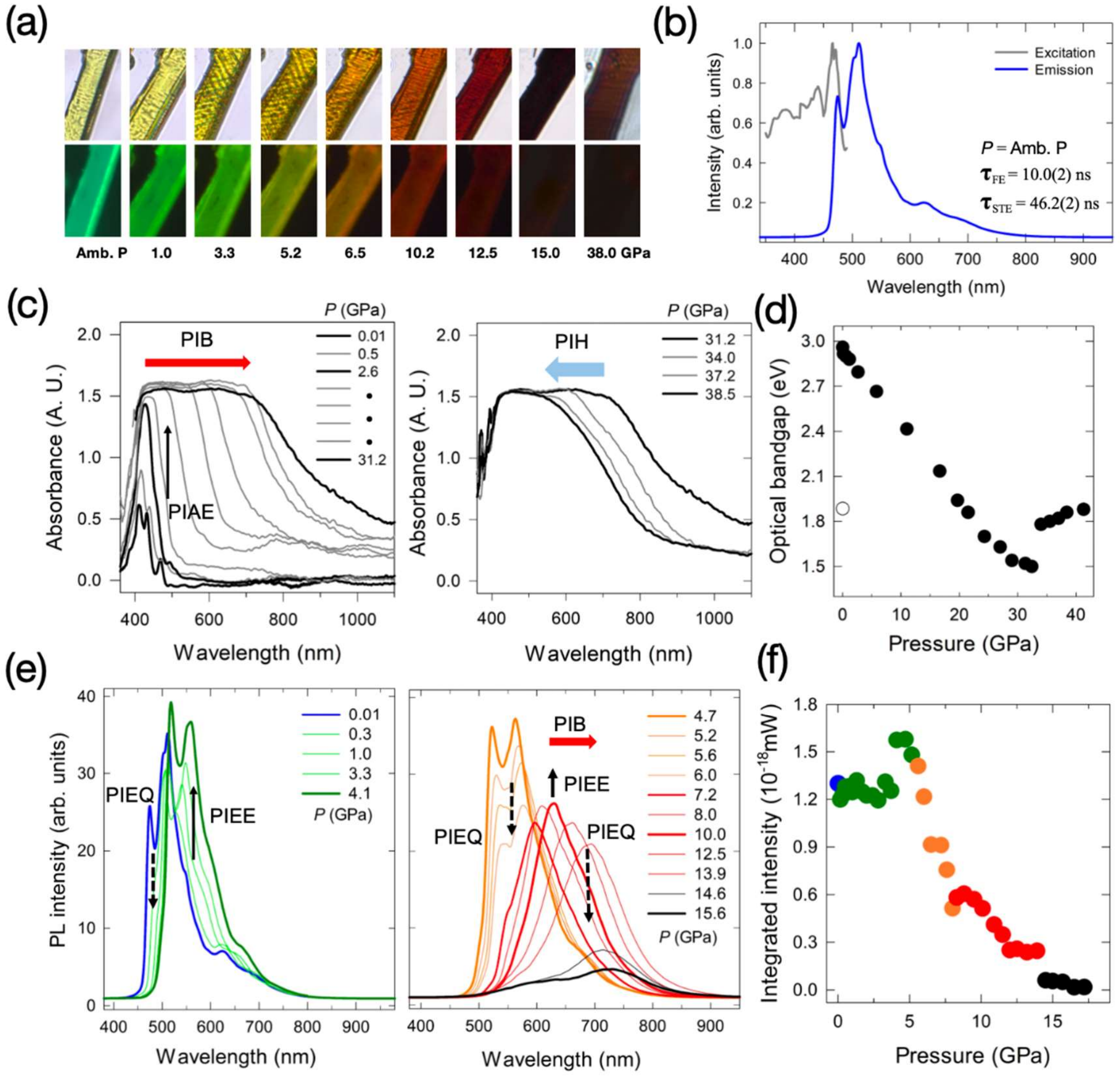

**Figure 4. Pressure-dependent optical properties of coronene.** (a) Optical and PL microphotographs of a coronene single-crystal compressed up to 38.0 GPa, illustrating a broad range of color tuneability. (b) Excitation (grey) and emission spectra (blue) at ambient conditions. (c) Pressure-dependent absorption spectra in absorbance unit showing PIAE effect up to 2.6 GPa and PIB effect up to 31.2 GPa (left), followed by the PIH effect above 31.2 GPa (right). (d) Estimated bandgap energy during compression (filled symbols) and decompression (open symbols). (e) Pressure-dependent PL spectra showing multiple regimes of PIEQ and PIEE effects. (f) Total integrated PL intensity as a function of pressure.

The optical images reveal a pronounced pressure-dependent color change, from pale-yellow (0.0001 GPa) to deep yellow (2.0 GPa), orange (6.5 GPa), red (12.5 GPa), dark red (14.0 GPa) and reddish-brown (35.0 GPa). The PL microphotographs likewise reveal systematic changes in emission color upon compression, evolving from green (0.0001 GPa) to yellow (3.6 GPa), orange (6.4 GPa), and red (12.5 GPa) (Figure 4a). Figure 4b presents the excitation and emission spectra of the coronene single-crystal under ambient conditions. The excitation and emission maxima were observed at 465 and 510 nm, respectively, with a relatively small Stokes shift. The PL spectrum spans approximately 450 - 750 nm and comprises two characteristic components: a sharp free-exciton (FE) emission band centered at $\lambda_{em}$ = 475 nm and multiple broad self-trapped-exciton (STE) emission bands centered at $\lambda_{em}$ = 510 nm. These features are consistent with previous reports.[34,35] The presence of the two emission components is further supported by time-resolved PL measurements, which reveal two distinct decay components with different lifetimes, assigned to FE and STE emission, respectively (Figure S12). The PL quantum yield (PLQY) of the coronene crystal was measured to be $\Phi_f = 0.153$, using $Al_2O_3$ as a reference. This value is approximately 15 times higher than that of coronene molecules in tetrahydrofuran (THF; $\Phi_f = 0.01$),[34] indicating enhanced emission in the crystalline state, consistent with aggregation-induced emission enhancement (AIEE). Under ambient condition, the Commission Internationale de l'Eclairage (CIE) chromaticity coordinates of the PL emission are ($x$, $y$) = (0.25, 0.48). The pronounced pressure-dependent color changes indicate substantial modifications to the optical response of coronene under compression. Figure 4c shows the in situ UV-vis-near-infrared (NIR) absorption spectra. Upon compression to 2.6 GPa, a pressure-induced absorption enhancement (PIAE) was observed. Further compression led to progressive broadening and a shift of the absorption edge toward longer wavelengths, indicating pressure-induced bathochromism (PIB) up to 30.0 GPa, with absorption extended across the visible (500-700 nm) and NIR (700-1000 nm) regions. At 31.0 GPa, the absorption edge shifted toward higher energies, corresponding to pressure-induced hypochromism (PIH).

The optical band gap was estimated from Tauc plots constructed from the absorption spectra[36] collected at each pressure (Figure S13) and plotted as a function of pressure (Figure 4d). The estimated band gap decreased monotonically from 2.94 eV at amb. P to 1.44 eV at 30.0 GPa. Above 31.0 GPa, however, this trend reversed, and the band gap increased, reaching 1.89 eV at 41.1 GPa. Upon decompression from 41.1 GPa to amb. P, the original absorption features were not recovered (Figure S14). This irreversible change suggests that compression above 30 GPa may induce a persistent structural or chemical modification in coronene, although further characterization is required to distinguish between these possibilities.

The pressure-dependent evolution of the PL emission provides further insight into the optical response of coronene. At 0.5 GPa, the integrated intensity of the FE emission band decreased by 85.3%, corresponding to pronounced pressure-induced emission quenching (PIEQ), whereas the integrated intensity of the STE emission increased by 23.7%, indicating pressure-induced emission enhancement (PIEE), relative to the amb. P. values. The suppression of the sharp FE emission, corresponding to a reduction in the blue component of the spectrum, resulted in a substantial shift in the CIE chromaticity coordinates from ($x$, $y$): (0.25, 0.48) to (0.29, 0.54) (Table S11, Figure S15). Interestingly, upon further compression up to 4.0 GPa, the total integrated PL intensity remained nearly constant, reflecting the approximate compensation between FE quenching and STE enhancement over this pressure range. Above 4.0 GPa, the FE emission disappeared, whereas the STE emission continued to increase, resulting in a maximum in the total integrated PL intensity at 4.7 GPa with approximately 35.8% higher that at 0.5 GPa. Above 5.2 GPa, the emission intensity began to decrease, while the emission spectrum continued to exhibit a redshift consistent with the PIB response. At 8.8 GPa, a second enhancement in PL intensity was observed, after which the emission became barely detectable at 17.5 GPa (Figure 4e). Notably, the original emission features were recovered upon decompression from 19.3 GPa to amb. P., whereas decompression from pressures above 30 GPa did not restore the original low-pressure emission spectrum (Figure S16). This difference indicates that the optical changes at moderate pressures are largely reversible, whereas compression beyond 30 GPa induces a persistent modification of the material.

### 2.3. Molecular deformation and intermolecular coupling reshape excited-state landscape

Unlike the simple emission quenching commonly reported for compressed organic molecular solids,[37] coronene exhibits a complex, nonmonotonic piezochromic response involving competing FE and STE emission processes. This behavior suggests that the optical response arises from the interplay between pressure-dependent intermolecular electronic coupling and intramolecular structural deformation.[38] At low pressures, the γ-to-β polymorphic transition substantially reorganizes the crystal packing, changing both the intermolecular geometry and the degree of π-surface overlap. This structural rearrangement coincides with pronounced suppression of the FE emission and enhancement of the STE emission. Upon compression, the distance between the adjacent coronene columns packed along the *b*-axis becomes closer along the *c*-axis causing structural distortion of each coronene, resulted in significant deviation from the original planarity ((179.7(3)°) **(Figure 5a)**. To further elucidate the role of molecular deformation, single-point calculations were performed on isolated gas-phase coronene molecules using geometries extracted from the crystal structures at each pressure points. The aromaticity indices, $NICS(1)_{zz}$ (nucleus-independent chemical shifts, HF/6-311G(d,p)), of rings A to D were suggested to be independent of pressure between 1.0 to 19.5 GPa, while only at the time when the pressure varies from 0.0 to 1.0 GPa, aromaticities of rings B, C, and D were weakened, which might contribute to the destabilization of coronene itself (Figure 5c). Changes in the Gibbs free energy at 298 K relative to the ambient-condition geometry were subsequently estimated at the B3LYP-D3/6-31G(d) level of theory (Figure 5d). Soon after the destabilization by $\Delta G$ +10.8 kcal/mol at 1.0 GPa where the γ- to β-phase transition occurs, the relative energy remains nearly unchanged (+3.5 to +5.2 kcal/mol) between 1.7--6.5 GPa upon focusing on a single molecule of coronene. A large energy leap from +4.5 to +18.6 kcal/mol was found again at 6.5--10.4 GPa and the energy plot approaches the least stable point (+19.5 kcal/mol) at 14.8 GPa. In ideal $D_{6h}$ symmetry, the HOMO/HOMO-1 and LUMO /LUMO+1 orbitals are doubly degenerate. Pressure-induced molecular deformation lowers the symmetry, lifting this orbital degeneracy and splitting the energy levels.

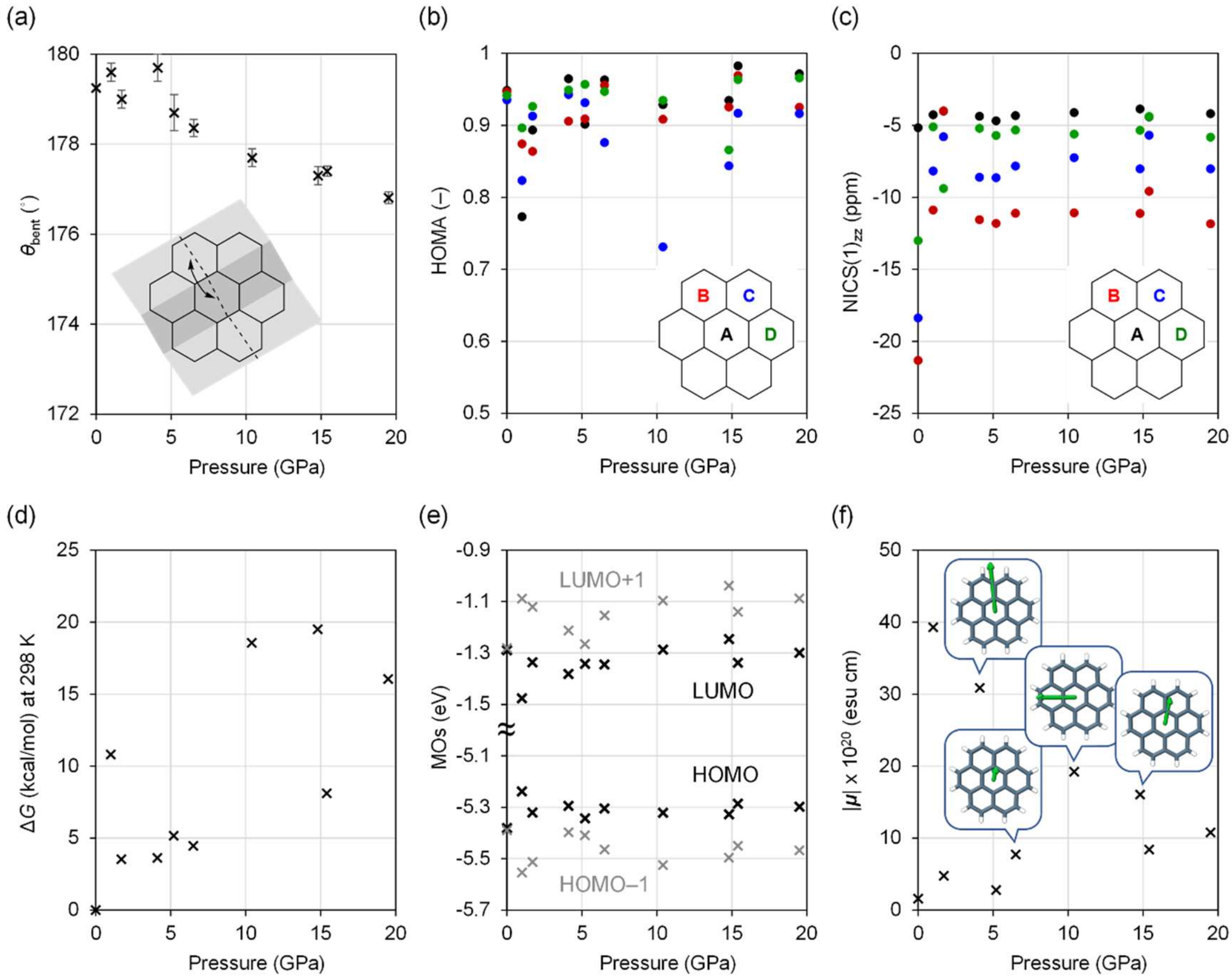

**Figure 5. Pressure-dependent molecular properties of coronene derived from crystallographic geometries.** (a) Molecular bent angles ($\theta_{bent}$). (b) Harmonic oscillator model of aromaticity (HOMA) indices. (c) NICS(1)$_{zz}$ indices calculated at the HF/6-311G(d,p) level. (d) Relative Gibbs energies ($\Delta G$) at 298 K and (e) molecular orbital (MO) energy-level diagrams calculated at the B3LYP-D3/6-31(G). (f) magnitude and arrangement of transition electric dipole moments ($|\boldsymbol{\mu}|$) for the $S_1 \rightarrow S_0$ excitation (TD CAM-B3LYP-D3/6-31(G)). Calculations in panels (c-f) were performed on a single molecule of coronene in gas-phase using the geometries extracted from the crystal structures at the corresponding pressures.

Finally, we examined how pressure-induced molecular deformation affects fluorescence. Given the relatively small Stokes shift (Figure 4b), structural relaxation in the $S_1$ excited state was assumed to be limited. Under this approximation, transition electric dipole moments ($|\mu|$) for the $S_1 \rightarrow S_0$ transition was calculated at the TD-CAM-B3LYP-D3/6-31G(d) level using molecular geometries extracted from the crystal structures (Figure 5f). The squared magnitude, $|\mu|^2$, was used as a measure of the relative transition probability. The largest calculated $|\boldsymbol{\mu}|$, 39.3 × $10^{-20}$ esu cm, occurs at 1.0 GPa, coinciding with the strong emission observed experimentally (Figure 4e). Following a decrease at 1.7 GPa, $|\boldsymbol{\mu}|$ rises again to 30.9 × $10^{-20}$ esu cm, corresponding to the first regime of PIEE. A similar increase, from 7.72 × $10^{-20}$ to 19.2 × $10^{-20}$ esu cm between 6.5 and 10.4 GPa, coincides with the second PIEE regime. At 10.4 GPa, the transition dipole orientation changes markedly, reflecting changes in the molecular orbital distributions. Further compression progressively reduces $|\boldsymbol{\mu}|$, consistent with the experimentally observed decline in PL intensity (Figure 4f). Beyond the first and second PIEE regimes, the PL intensity decreases above approximately 5.2 and 10.4 GPa, respectively. These decreases coincide with changes in intramolecular structural descriptors, including variations in bond lengths (Figure S17 and Table S11). To quantify the associated changes in bond length, we calculated the harmonic oscillator model of aromaticity (HOMA) indices for rings A-D (Figure 5b). At ambient pressure (0.0001 GPa), all four rings exhibit HOMA values close to unity, reflecting the high symmetry of coronene. For ring C, HOMA decreases from 0.876(38) at 6.5 GPa to 0.731(67) at 10.4 GPa, although this difference is not statistically significant within the reported uncertainties, consequently leading to almost identical aromaticity indices (Figure 5d). The lower HOMA value at 10.4 GPa is largely attributed to a K-region bond in ring C, whose refined length of 1.290(17) Å is much shorter than the ideal benzene reference value of 1.397 Å.

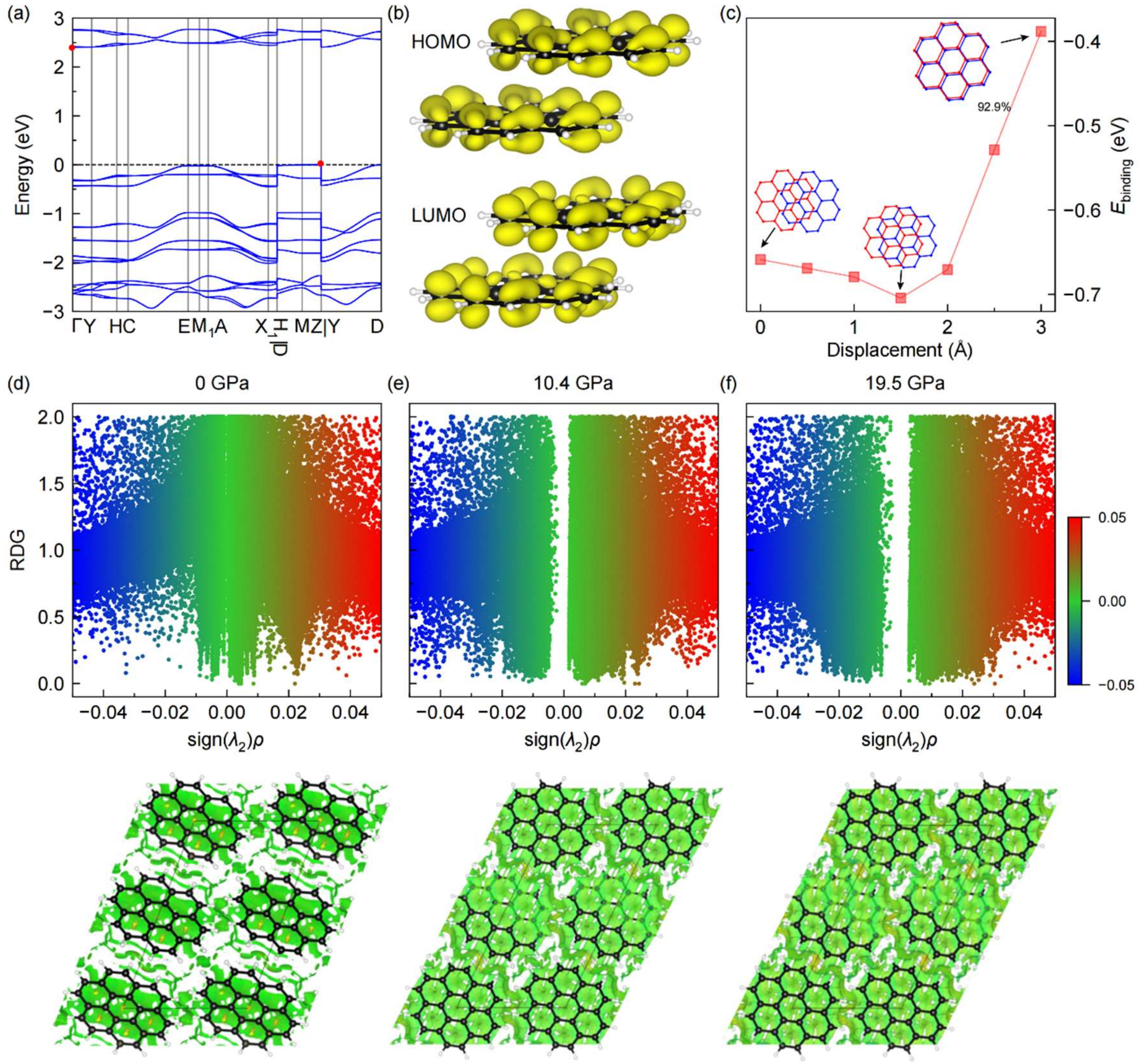

**Figure 6. Electronic structure and intermolecular interactions in crystalline coronene.** (a) Electronic structure at 0 GPa, with red points marking the HOMO and LUMO positions. (b) Charge-density distributions of the HOMO (top) and LUMO (bottom), plotted at isosurface value of 0.01 e/Å$^3$. (c) Binding energy ($E_{binding}$) as a function of the relative lateral displacement between the upper and lower coronene

molecules, defined as $E_{binding} = E_{\text{total}} – E_{\text{upper}} – E_{\text{lower}}$, where $E_{\text{total}}$ is the total energy of the molecular pair and $E_{\text{upper}}$ and $E_{\text{lower}}$ are the energies of the corresponding isolated molecules. (d–f) Reduced density gradient (RDG) plots as a function of sign($\lambda_2$)ρ at 0, 10.4, and 19.5 GPa, respectively. The lower panels show the corresponding noncovalent interaction (NCI) isosurfaces at RDG = 0.5, colored according to sign($\lambda_2$)ρ. Blue, green and red regions correspond to attractive, weak van der Waals, and repulsive interactions, respectively.

To further elucidate intermolecular interactions under pressure, we performed DFT calculations. At 0 GPa, crystalline coronene exhibits an insulating electronic structure with a calculated bandgap of 2.4 eV **(Figure 6a)**. Although this value differs from the experimentally estimated optical bandgap, the calculations qualitatively reproduce the insulating character observed experimentally (Figure 4d). The HOMO and LUMO are dominated by out-of-plane C $p_z$ orbitals that form π-conjugated states (Figure 6b), consistent with intermolecular π–π coupling between adjacent molecules in the face-to-face arrangement. Pressure modifies these interactions by changing both the intermolecular separation and the relative stacking geometry of adjacent coronene molecules. To isolate the effect of lateral displacement, we calculated the binding energy ($E_{\text{binding}}$) for a pair of molecules shifted rigidly relative to one another, thereby varying their π–π overlap. As shown in Figure 6c, $E_{\text{binding}}$ decreases with increasing displacement $d$, reaching a minimum at approximately 1.5 Å in an AB-like configuration, before increasing toward an AA-like configuration. This increase is attributed to enhanced Pauli repulsion at maximal π-orbital overlap. These results suggest that pressure-induced sliding toward an AB-like arrangement favors intermolecular stabilization. Consistent with this interpretation, reduced density gradient (RDG) analysis reveals a depletion of features near $\text{sign}(\lambda_2)\rho \approx 0$and a redistribution toward both attractive and repulsive regions with increasing pressure (Figures 6d–f), indicating an evolving balance of intermolecular interactions. The corresponding NCI maps show increasingly extensive green isosurfaces between neighboring molecules, consistent with a larger intermolecular contact region under compression.

Based on the combined experimental and DFT results, we propose the following pressure-dependent mechanism for the optical response of coronene. Under ambient conditions, coronene exhibits dual fluorescence, FE and STE emissions, within its γ-herringbone packing. Upon compression to approximately 2.0 GPa, the FE emission is suppressed, coinciding with the phase transition from γ- to β-$C_{24}H_{12}$. Meanwhile, the gradual enhancement of STE emission above 1.0 GPa is tentatively attributed to pressure-induced molecular deformation, as evidenced by SC-XRD and Raman measurements. In flexible organic molecules, excitons can become self-trapped through exciton-induced molecular relaxation, including twisting or bending. Such structural relaxation can strengthen exciton-phonon coupling and stabilize self-trapped excited states, thereby promoting radiative STE recombination. In coronene, pressure-induced molecular pre-deformation becomes apparent above approximately 1.2 GPa, accompanying a 35.8% increase in PL intensity up to approximately 5 GPa. Above 5 GPa, a decrease in the calculated transition dipole moments is observed. In this pressure regime, the interplanar distance decreases to below 3.2 Å, potentially strengthening intermolecular interactions and suppressing STE emission by increasing exciton delocalization across neighboring molecules. Consequently, the PL intensity continues to decrease until approximately 8 GPa. Above 8.0 GPa, a second PIEE, accompanied by PL peak broadening and pronounced PIB, is observed. This response coincides with a reduction in the interplanar separation to below 3.0 Å, an increase in the π-π overlap ratio to 72.9%, deformation of ring C, and a change in the calculated relative Gibbs free energy of the isolated molecule. Together, these observations suggest that enhanced intermolecular electronic coupling may contribute to the increase transition dipole moments and the possible formation of excimer-like states in the 8.0-10.4 GPa range. At approximately 15 GPa, the total PL intensity decreases markedly, potentially reflecting increased structural fluctuations involving ring A. Optical measurements further indicate an irreversible transformation following compression above 30 GPa. Although structural data are unavailable beyond 19.5 GPa, extrapolation of the EoS data suggests that the interplanar separation within the π-stacked arrangement could approach 2.7 Å at 30 GPa. Such close intermolecular contact may facilitate intermolecular bonding or cross-linking, although this interpretation remains speculative without direct structural evidence. Notably, the IR spectrum of the recovered product resembles that of amorphous hydrogenated carbon produced by compressing benzene,[39, 40] providing additional support for an irreversible transformation of coronene under extreme pressure (Figure S18).

## 3. Conclusion

This work establishes crystalline coronene as a platform for resolving how molecular packing and deformation jointly govern the optical response of stimuli-responsive molecular solids. By integrating high-pressure SC-XRD, Raman spectroscopy, optical absorption and photoluminescence measurements with DFT calculations, we distinguish the structural reorganization associated with the γ-to-β polymorphic transition from the subsequent progressive deformation of an initially planar π-conjugated framework. These successive structural regimes accompany a nonmonotonic optical response, encompassing redistribution between FE and STE emission, alternating photoluminescence enhancement and quenching, and pronounced spectral shifts.

The combined results reveal a pressure-dependent interplay between intramolecular distortion and intermolecular coupling – a relationship between pressure-induced molecular deformation and the optical response of a prototypical planar PAH. Calculations identify changes in aromaticity-related descriptors, molecular energetics, and transition dipole moments that help rationalize the observed emission behavior, while increasingly close π-stacking introduces additional pathways for intermolecular electronic interactions. Together, these findings provide an experimentally grounded framework for understanding how deviations from molecular planarity and changes in crystal packing reshape excited-state behavior.

Beyond coronene, this study highlights molecular deformation as a design variable complementary to chemical modification and packing control. Hydrostatic pressure thus serves not only as a means of tuning optical properties but also as a mechanistic probe that connects molecular-scale structural evolution with macroscopic photophysical function, opening opportunities for the rational design of stimuli responsive π-conjugated materials.

[Deposition numbers CCDC 2374515 – 2374517 and 2564243 – 2564251 contain the supplementary crystallographic data for this paper. These data can be obtained free of charge via www.ccdc.cam.ac.uk/data_request/cif, by emailing data_request@ccdc.cam.ac.uk, or by contacting The

Cambridge Crystallographic Data Centre, 12 Union Road, Cambridge CB2 1EZ, UK; fax: +44 1223 336033.]

**Acknowledgements**

The manuscript was written through contributions of all authors. All authors have given approval to the final version of the manuscript. These authors (W.Z. and S.Y.) contributed equally. The authors acknowledge the Diamond Light Source (DLS) for provision of beamtime at the I15 beamline (proposal No. CY34815-1 and CY39434-1), National Synchrotron Radiation Research Center (NSRRC) for beamtime at SPring-8, BL12B2 (proposal No. 2023A4139 and 2024B4138), and the High Energy Photon Source (HEPS) ID31 High Pressure Beamline (https://cstr.cn/31138.02.HEPS.ID31) for providing technical support and assistance in XRD data collection. For the purpose of open access, the author has applied a Creative Commons Attribution (CC BY) license to any Author Accepted Manuscript version arising from this submission.

**Funding:** T.N. gratefully acknowledges funding from the Beijing Natural Science Foundation (Project No. IS24025). D.L. and A.L. thank the UKRI Future Leaders Fellowship (MR/V025724/1) for financial support. P. D-S. acknowledges the support of the National Natural Science Foundation of China (NSFC) reference code W2532012. Y.H. acknowledges financial support from JACI Prize for Encouraging Young Researcher, Yazaki Memorial Foundation for Science and Technology, the Asahi Glass Foundation, Kurita Water and Environment Foundation, Shitagau Noguchi Research Grant (NJ202504), The Samco Foundation, and The Mitsubishi Foundation (202611031). Y.D. thank the National Key Research and Development program of China (Grant No. 2022YFA1402301). H.G. thank the National Science Fund for Distinguished Young Scholars (Grant No. T2225027).